\documentclass[10pt]{iopart}
\pdfoutput=1

\usepackage{graphicx}
\usepackage{color}
\usepackage{iopams}  

\begin{document}

\title{On the relation between uncertainties of weighted frequency averages and the various types of Allan deviations}
\author{Erik Benkler, Christian Lisdat, Uwe Sterr}
\ead{erik.benkler@ptb.de}

\address{Physikalisch-Technische Bundesanstalt, Bundesallee 100, 38116 Braunschweig, Germany}

\vspace{10pt}
\begin{indented}
\item[]\today
\end{indented}

\begin{abstract}
The power spectral density in Fourier frequency domain, and the different variants of the Allan deviation (ADEV) in dependence on the averaging time are well established tools to analyse the fluctuation properties and frequency instability of an oscillatory signal. It is often supposed that the statistical uncertainty of a measured average frequency is given by the ADEV at a well considered averaging time. However, this approach requires further mathematical justification and refinement, which has already been done regarding the original ADEV for certain noise types. Here we provide the necessary background to use the modified Allan deviation (modADEV) and other two-sample deviations to determine the uncertainty of weighted frequency averages. The type of two-sample deviation used to determine the uncertainty depends on the method used for determination of the average. We find that the modADEV, which is connected with $\Lambda$-weighted averaging, and the two sample deviation associated to a linear phase regression weighting (parADEV) are in particular advantageous for measurements, in which white phase noise is dominating. Furthermore, we derive a procedure how to minimize the uncertainty of a measurement for a typical combination of white phase and frequency noise by adaptive averaging of the data set with different weighting functions. Finally, some aspects of the theoretical considerations for real-world frequency measurement equipment are discussed.
\end{abstract}

%
%
%
%
\ioptwocol
%

%
%
%
%
%
%
\section{Introduction} \label{sec:intro}
The central outcome of a frequency measurement is an average frequency and its associated uncertainty. The determination of the uncertainty depends on the method to determine the average frequency. The topic of this paper is the stringent derivation of the relations between averaging methods and associated uncertainty, and the identification of averaging methods, which are advantageous in the sense of approaching the tightest uncertainty for given noise types.

The average frequency value determined in a single measurement can only be an estimate of the true frequency, because even if there are no unknown systematic offsets, we only have limited knowledge about the actual value due to noise which always contributes to a measurement. Often, the term ``statistical uncertainty'' is used with this meaning, and in the following, we will use the term ``uncertainty'' or ``$u$'' as a short form.  

Traditionally, frequency has been most often measured with dead-time free so called $\Pi$-counters. The most natural measure of frequency instability determined from $\Pi$-counted frequency values is the Allan deviation (ADEV)~\cite{all66, all87}. Besides being a measure for frequency instability, the ADEV is a well established tool to derive information about the type of noise that is present in the measurement, e.g.~\cite{bar71,les84,rub05,daw07}. The ADEV shows a $\tau^{-1/2}$ averaging behaviour for white frequency noise, i.e.\ for uncorrelated frequency values, and a steeper $\tau^{-1}$ power law for white phase noise, due to correlations between the according frequency values in the latter case. 

The question if an estimate of the ADEV $\sigma_y(\tau=T)$ at the total length $T$ of a measurement can be used for the determination of the uncertainty $u$ has been addressed in~\cite{lee10}. There it is shown that the ADEV is equal to $u$ for white frequency noise, while for white phase noise a constant correction factor of order unity has to be applied. It is thus justified to profit from the fast reduction of $\sigma_y(\tau)$ with $\tau^{-1}$ in the case of white phase noise. 

However, the $\Pi$-counter approach is not the only choice to measure frequencies, and in fact not always the optimum one. Especially the fact that $\Pi$-counters only use two phase measurements -- essentially detecting the times of the signal's zero-crossings at the beginning and at the end of the measurement interval -- makes this technique very susceptible to wide bandwidth phase noise. 

For this reason, counters have been developed that average the phase over parts of the interval, in order to effectively reduce the bandwidth and thus the influence of phase noise. In the extreme case, for an averaging interval with length $\tau$ (actually data over an interval of at least $2\tau$ is required), the phase is averaged between times $0 \dots \tau$ and $\tau \dots 2\tau$, which would be called a $\Lambda$-counter. As we will show, the appropriate method to determine uncertainties associated to $\Lambda$-weighted averages is the modified Allan deviation ${\rm mod} \, \sigma_y (\tau)$. 

Practically, ${\rm mod} \, \sigma_y (\tau)$ has been introduced to resolve the ambiguity that is present in the functional dependence of the ADEV on $\tau$ between white and flicker phase noise and to avoid a dependence of the ADEV on the measurement bandwidth in these cases~\cite{all81}. As the modADEV shows an even steeper $\tau^{-3/2}$-power law for white phase noise than the ADEV, the question arises if $\Lambda$-averaging could be employed advantageously to achieve a reduced uncertainty $u$ by using the associated modADEV for its determination. 

In many cases, white frequency noise dominates at long averaging times, and thus a rejection of phase noise by the modADEV is not necessary~\cite{rub05}. As we will show, the use of the modADEV is even slightly detrimental in this case.
There are, however, important measurement examples, for which white phase noise dominates even for $\tau = 10^3\,\mathrm{s}\dots 10^6\,\mathrm{s}$~\cite{rau15,rau14,pre12,bau06}, and the uncertainty $u$ can be reduced significantly by a better data averaging and uncertainty analysis strategy, as for example by $\Lambda$-averaging in conjunction with uncertainty determination employing the modADEV.

In section~\ref{sec:modADEV}, the relationships between the uncertainty of $\Pi$- and $\Lambda$-weighted averages of a measurement data set and  the ADEV and modADEV are derived, since these are the technically widely implemented types. In section~\ref{sec:noise}, the conversion factors from the modADEV to the uncertainty of the corresponding $\Lambda$-weighted average are deduced for different noise types. The question of an optimum averaging method is addressed in section~\ref{sec:optimum}, whereas the topic is discussed in terms of correlations in section~\ref{sec:cor}. In section~\ref{sec:mixed}, a strategy for the transition between white phase and white frequency noise is povided. Finally, some practical issues are discussed in section~\ref{sec:practical}.
%
%
%
%
%
%
\section{Average frequency and associated uncertainty} \label{sec:modADEV}
The value of an oscillatory signal with amplitude $V_0$, phase $\Phi(t)$ and nominal frequency $\nu_0$ is described by 
\begin{equation}
V(t)=V_0\sin \left[\Phi(t)\right]=V_0\sin \left[2 \pi \nu_0 t +\phi(t)\right].
\label{eq:phase}
\end{equation}
The residual phase $\phi(t)$ contains the fluctuations due to noise and thus the statistical properties of the signal, whereas the nominal frequency $\nu_0$ is defined as a time-independent value near the ``true average'' of the oscillator frequency. This true frequency cannot be perfectly known as a matter of principle, but only be estimated by performing a finite number of experiments. The instantaneous frequency is defined by
\begin{equation}
\nu(t) \equiv \frac{1}{2 \pi}\frac{\rmd \Phi}{\rmd t} = \nu_0 + \frac{1}{2 \pi}\frac{\rmd \phi}{\rmd t}.
\label{eq:freq}
\end{equation}
Usually, the fractional frequency deviation $y(t)$ is defined as
\begin{equation}
y(t) \equiv \frac{\nu(t)}{\nu_0} - 1 =\frac{1}{2 \pi \nu_0}\frac{\rmd \phi}{\rmd t}.
\label{eq:frac_freq}
\end{equation}
The targeted result of a frequency measurement is an estimator $\overline{y}$ for the expectation value $\mu=\langle y \rangle$ of the fractional frequency, along with a suitable uncertainty $u(\overline{y})$. The expectation value is defined as the average over all elements of a fictitious ensemble of an infinite number of independent realizations of the random variable $y$. Throughout this paper, the angled brackets indicate the ensemble average.
The uncertainty is defined as the scatter of the weighted averages $\overline{y}_w$ around the expectation value $\mu$, where the $\overline{y}_w$ are the realizations mentioned above. This scatter is described by the classical standard deviation, i.e. the squared uncertainty is defined by the classical variance
\begin{equation}
u^2 = \left\langle (\overline{y}_w - \mu)^2  \right\rangle = \left\langle (\overline{y}_w - \left\langle \overline{y}_w \right\rangle)^2  \right\rangle.
\label{eq:classical}
\end{equation}
As mentioned in the introduction, this ``statistical'' uncertainty does not include other contributions to the overall uncertainty besides those from noise. To determine $u(\overline{y})$, the classical variance can be estimated by analysing the noise properties of the underlying measurement data. In the frequency domain, the power spectral density $S_y(f)$ contains the necessary information about the fluctuations, whereas in the time domain the Allan variance $\sigma_y^2 (\tau)$ and its derived variants are usually employed to analyse the noise properties.

For convenience of reading, we will from here on not distinguish explicitly between the true uncertainty given by the classical variance and its estimate. This does not change the presented results for practical purposes. Furthermore, we presuppose without further discussion at several steps of the mathematical derivations that the underlying noise processes are ergodic.

In the following, we will first treat the estimation of the expectation value $\mu$ using $\Pi$- and $\Lambda$-weighting as prominent examples of averaging methods. In a second step, we derive equations relating the power spectral density with the uncertainty $u$ associated to the averaging method. In a third step, we derive equations relating the power spectral density with the Allan variance variants corresponding to the respective averaging methods. Finally, we compare the equations obtained in the second and third step, in order to find the noise conditions under which the uncertainty $u$ can be expressed using the according Allan deviation, and to find the appropriate conversion factors between them. 

\begin{figure}
\centerline{\includegraphics[width=1\columnwidth]{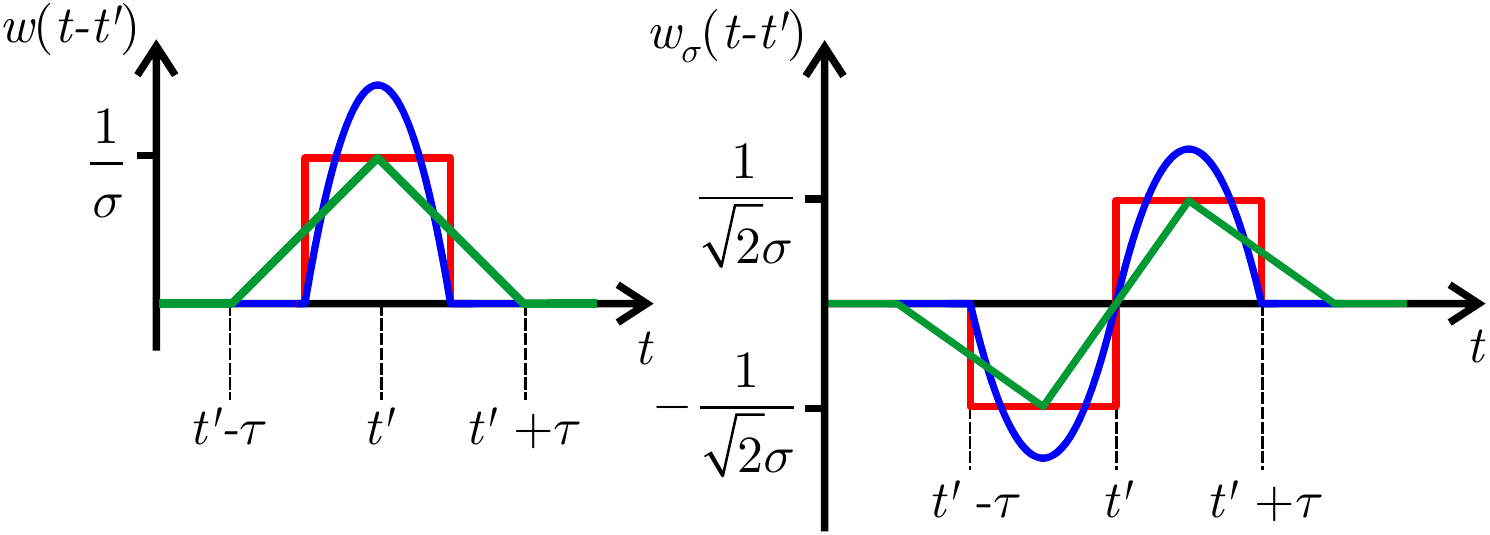}}
\caption{Weighting functions for $\Pi$- (red), $\Lambda$- (green) and $\Omega$- (blue) counters (left), and for the corresponding Allan deviation types (right).}
\label{fig:sensfct}
\end{figure}

We describe an averaging method by means of a weighting function $w(t)$, which is non-zero only over a finite time interval between $t=-\tau/2$ and $t=\tau/2$. Furthermore, we define a continuous moving weighted average function $\overline{y}_w(t)$ given by the convolution between the instantaneous fractional frequency $y(t)$ and the weighting function. $\overline{y}_w(t)$ describes a sample of $y(t)$ averaged over an interval of length $\tau$ around $t$
\begin{equation}
\overline{y}_w(t) = \int^{\infty}_{-\infty}{y(t^\prime)w(t^\prime - t)\, \rmd t^\prime}.
\label{eq:y_w}
\end{equation}
We see from equations~\ref{eq:classical} and~\ref{eq:y_w} that $u(\overline{y})$ is intrinsically connected to the weighting function, i.e.\ to the method used for the determination of the average value. Different weighting functions are known, e.g.\ for rectangularly weighted, i.e\ $\Pi$-averaging as with an ideal $\Pi$-counter
\begin{equation}
w_\Pi(t) = \left\{  \begin{array}{ll} 
				1/\tau 		&  -\tau/2 < t \leq \tau/2 \\
				0  				& {\rm elsewhere} 
																	\end{array}, \right.
\label{eq:sens_av_pi}
\end{equation}
and for $\Lambda$-averaging
\begin{equation}
w_\Lambda(t) = \left\{  \begin{array}{ll} 
				1+t/\tau^2 		& -\tau < t \leq 0 \\
				1-t/\tau^2	&	 0 < t \leq \tau \\
				0  				& {\rm elsewhere} 
																	\end{array}. \right.
\label{eq:sens_av_lam}
\end{equation}
Another interesting averaging method is to perform a least-squares fit of a linear function to the phase $\phi(t)$ over the interval $-\tau/2\dots \tau/2$~\cite{joh05,daw07}. According to equation~\ref{eq:frac_freq}, the average fractional frequency $\overline{y}_w$ at $t=0$ is given by the slope of the regression line divided by $2\pi\nu_0$. For the regression,
\begin{equation}
\chi^2=\int_{-\tau/2}^{\tau/2} \left[\phi(t)-\left(\phi_0 + 2\pi\nu_0\overline{y}_w t\right)\right]^2 \, \rmd t
\end{equation}
is minimized with respect to the parameters $\phi_0$ and $\overline{y}_w$ of the regression line. Hence, $\frac{\partial\chi^2}{\partial\phi_0}=0$ and 
\begin{equation}
\frac{\partial\chi^2}{\partial\overline{y}_w}=-4\pi\nu_0\int_{-\tau/2}^{\tau/2} t \phi(t) - t\phi_0-2\pi\nu_0\overline{y}_wt^2 \, \rmd t = 0,
\label{eq:int_avg_angfreq}
\end{equation}
which implies
\begin{eqnarray}
\frac{\overline{y}_w\tau^3}{12} &=& \frac{1}{2\pi\nu_0}\int_{-\tau/2}^{\tau/2} t\phi(t)\, \rmd t \nonumber\\
&=&\frac{1}{2\pi\nu_0}\left[\frac{\tau^2}{8}\int_{-\tau/2}^{\tau/2} \frac{\rmd \phi}{\rmd t} \, \rmd t- \int_{-\tau/2}^{\tau/2} \frac{t^2}{2} \frac{\rmd \phi}{\rmd t} \, \rmd t \right],
\label{eq:int_omega_phi}
\end{eqnarray}
where we have integrated by parts. Hence with equation~\ref{eq:frac_freq} 
\begin{eqnarray}
\overline{y}_w(t)&=& \int_{-\tau/2}^{\tau/2} \left(\frac{3}{2\tau}-\frac{6t^2}{\tau^3}\right) y(t)\, \rmd t \nonumber\\
&=& \int_{-\tau/2}^{\tau/2} w_\mathrm{par}(t)y(t)\, \rmd t
\label{eq:omega}
\end{eqnarray}
and the frequency weighting function for linear regression of phase data is found as
\begin{equation}
w_\mathrm{\Omega}(t)= \left\{  \begin{array}{ll} 
				3/(2\tau)-6t^2/\tau^3	&  -\tau/2 < t \leq \tau/2 \\
				0  				& {\rm elsewhere} 
																	\end{array}. \right.
\label{eq:w_reg}
\end{equation}
In analogy to the well-known $\Pi$- and $\Lambda$-counters and corresponding weighting functions, the nomenclature $\Omega$-counter and  $\Omega$-weighting function is used due to the parabolic shape of the weighting function associated with a linear regression to phase data \cite{ben15a}. The temporal weighting functions are shown in figure~\ref{fig:sensfct}. 

Plugging equation~\ref{eq:y_w} 
into equation~\ref{eq:classical} yields for the uncertainty
\begin{equation}
u^2 = \left\langle \left[ \int^\infty_{-\infty} y(t^\prime) w(t^\prime - t) \, \rmd t^\prime -\mu\right]^2  \right\rangle.
\label{eq:u}
\end{equation}
Using Parseval's theorem to express this by the single-sided power spectral density and Fourier transform $W(f)=\int_{-\infty}^\infty \exp(2\pi i f t) w(t) \,\rmd t$ of the weighting function yields
\begin{equation}
u^2 = \int^\infty_0 S_y(f) \left| W(f) \right|^2 \,\rmd f
\label{eq:u_FT}
\end{equation}
with
\begin{equation}
W_{\Pi}(f) = \frac{ \sin(\pi f \tau)}{(\pi f \tau)},
\label{eq:FT_sens_Pi}
\end{equation}
\begin{equation}
W_{\Lambda}(f) =  \frac{ \sin^2(\pi f \tau)}{(\pi f \tau)^2}
\label{eq:FT_sens_Lam}
\end{equation}
and 
\begin{equation}
W_\Omega(f)=\frac{3\sin\left(\pi f \tau \right)}{(\pi f \tau)^3}-\frac{3\cos\left(\pi f \tau \right)}{(\pi f \tau)^2}.
\label{eq:w_reg_FT}
\end{equation}
The Allan variance is the two-sample variance without dead time
\begin{equation}
\sigma_y^2 (\tau) = \frac{1}{2}\left \langle \left[ y_{w_\Pi}(t+\tau/2) - y_{w_\Pi}(t-\tau/2) \right]^2  \right\rangle,
\label{eq:ADEV}
\end{equation}
where the two samples are the fractional frequency averages taken over two intervals of duration $\tau$ separated by $\tau$, i.e.\ without dead time. We now develop the mathematical justification and noise conditions for determining the uncertainty from the two-sample variance associated to a frequency averaging method described by a weighting function $w(t)$.

By inserting equations~\ref{eq:y_w} and \ref{eq:sens_av_pi} into equation~\ref{eq:ADEV}, the ADEV can be expressed as
\begin{equation}
\sigma_y^2 (\tau) = \left\langle \left[ \int^\infty_{-\infty} y(t^\prime) w_\sigma(t^\prime - t)\, \rmd t^\prime \right]^2  \right\rangle
\label{eq:ADEV2}
\end{equation}
with the weighting function for the ADEV given by
\begin{eqnarray}
w_\sigma (t) & = & \frac{1}{\sqrt{2}} \left[w_\Pi (t+\tau/2) - w_\Pi (t-\tau/2)\right]. 
\label{eq:sens_adev}
\end{eqnarray}
The definition of the ADEV in equation~\ref{eq:ADEV} shows the connection between the ADEV and $\Pi$-averaging: The Allan variance is defined as the two-sample variance of $\Pi$-averaged samples. This can be used to formulate a generalized definition of different types of Allan variances as the two-sample variance, where the two samples are determined using the same weighting function as it is used for the average value, whose uncertainty shall be determined. This generalized definition of the two-sample variance for a certain averaging type with xxx used as a placeholder for the type description reads as
\begin{eqnarray}
{\rm xxx} \, \sigma_y^2 (\tau) &=&  \frac{1}{2}\left\langle \left[ \overline{y}_{w_{\rm xxx}}(t+\tau/2) - \overline{y}_{w_{\rm xxx}}(t-\tau/2) \right]^2  \right\rangle \nonumber \\
&=&  \left\langle \left[ \int^\infty_{-\infty} y(t^\prime) w_{{\rm xxx} \, \sigma}(t^\prime - t)\, \rmd t^\prime \right]^2  \right\rangle,
\label{eq:modADEV}
\end{eqnarray}
with the one-sample weighting function $w_{\rm xxx}(t)$ and the two-sample weighting function $w_{{\rm xxx} \, \sigma}(t)$ associated to the averaging type xxx, given by
\begin{eqnarray}
w_{{\rm xxx} \, \sigma}(t) & = & \frac{1}{\sqrt{2}} \left[w_{\rm xxx} (t+\tau/2) - w_{\rm xxx} (t-\tau/2)\right].
\label{eq:sens_modadev}
\end{eqnarray}
For example, if the averaging type is $\Lambda$, the associated two-sample variance is the modADEV, and for phase regression averaging it is the parabolic Allan deviation, abbreviated here as parADEV~\cite{ben15a}.

Applying Parseval's theorem to equation~\ref{eq:modADEV}, the generalized Allan variance can be expressed by the single-sided power spectral density $S_y(f)$ and Fourier transform $W(f)$ of the temporal weighting function $w(t)$~\cite{ril08}. In the general case, the difference between the two samples time-shifted by $\tau$ as they are used in the two-sample variance leads to an interference term in Fourier space, i.e.\ to multiplication of the according spectral weighting function $W(f)$ by the factor $\sqrt{2}\sin(\pi f \tau)$ 
\begin{eqnarray}
{\rm xxx} \,\sigma_y^2 (\tau) &=& \int^\infty_0 S_y(f) \left| W_{{\rm xxx} \, \sigma} (f) \right|^2 \,\rmd f  \nonumber \\
&=& 2\int^\infty_0 S_y(f) \left| W_{\rm xxx}(f) \sin(\pi f \tau)\right|^2 \,\rmd f.
\label{eq:FT_modadev}
\end{eqnarray}
From this equation the well known power laws for the scaling of modADEV (and similarly ADEV) with averaging time $\tau$ can be calculated for different noise types and their corresponding power spectral densities $S_y$.

To establish the connection between the two-sample variance (e.g.\ modADEV or ADEV) and the uncertainty $u(\overline{y})$, equations~\ref{eq:FT_modadev} and \ref{eq:u_FT} are compared with each other, leading to the values detailed in the next section.
%
%
%
%
%
\section{Application to different noise types} \label{sec:noise}
\begin{table*}
\centering
\begin{tabular}{c|c|c|c|c|c}
\hline
\hline 
\rule[-2ex]{0pt}{5ex}	 & White PM	& Flicker PM	&	White FM & Flicker FM	& Random Walk FM\\ 
\hline
\rule[-2ex]{0pt}{5ex} $S_y(f)$ & $h_2 f^2$ & $h_1 f$ & $h_0$ & $h_{-1} f^{-1}$ & $h_{-2}f^{-2}$\\
\hline 
\rule[-4ex]{0pt}{7ex} $\sigma^2_y(\tau)$ & $\frac{3 f_h}{4 \pi^2} h_2 \tau^{-2}$	& $\frac{\ln(2\pi f_h\tau)+\gamma}{2\pi^2} h_1 \tau^{-2}$	& $\frac{1}{2} h_0 \tau^{-1}$	& $2 \ln (2) h_{-1}$ & $\frac{2}{3} \pi^2 h_{-2}\tau $\\
\rule[-2ex]{0pt}{5ex} $u_\Pi^2(\tau)$	& $\frac{f_h}{2 \pi^2} h_2 \tau^{-2}$	& $\frac{3\ln(2\pi f_h\tau)-\ln(2)+3\gamma}{4\pi^2} h_1 \tau^{-2}$ & $\frac{1}{2} h_0 \tau^{-1}$ & $\infty$ & $\infty$ \\
\rule[-2ex]{0pt}{4ex}	&	$= \frac{2}{3} \sigma^2_y(\tau)$& $\approx \frac{2}{3} \sigma^2_y(\tau)$	&	$= \sigma^2_y(\tau)$ &  &  \\
\hline 
\rule[-4ex]{0pt}{7ex} ${\rm mod} \, \sigma_y^2 (\tau)$& $\frac{3}{8 \pi^2} h_2 \tau^{-3}$	& $\frac{24\ln(2)-9\ln(3)}{8\pi^2} h_1 \tau^{-2}$	& $\frac{1}{4} h_0 \tau^{-1} $	& $\frac{27}{20} \ln (2) h_{-1}$	& $\frac{11}{20} \pi^2 h_{-2}\tau $	\\
\rule[-2ex]{0pt}{5ex} $u_\Lambda^2(\tau)$	& $\frac{1}{4 \pi^2} h_2 \tau^{-3}$	& $\frac{\ln (2)}{\pi^2} h_1 \tau^{-2}$ & $\frac{1}{3} h_0 \tau^{-1}$	& $\infty$ & $\infty$\\
\rule[-2ex]{0pt}{4ex}	&	$= \frac{2}{3} {\rm mod} \, \sigma_y^2(\tau) $	& $\approx 0.822 \,{\rm mod} \, \sigma_y^2(\tau) $ & $= \frac{4}{3} {\rm mod} \, \sigma_y^2(\tau) $ &  \\
\hline 
\rule[-4ex]{0pt}{7ex} $\mathrm{par}\,\sigma_y^2 (\tau)$& $\frac{3}{2 \pi^2} h_2 \tau^{-3}$	& $\frac{12\ln(2)-3}{2\pi^2} h_1 \tau^{-2}$	& $\frac{3}{5} h_0 \tau^{-1} $& $\frac{14-8\ln(2)}{5}h_{-1}$ & $\frac{26 }{35}\pi^2 h_{-2} \tau$  	\\
\rule[-2ex]{0pt}{5ex} $u_\Omega^2(\tau)$	& $\frac{3}{2 \pi^2} h_2  \tau^{-3}$	& $\frac{9}{4\pi^2} h_1 \tau^{-2}$ & $\frac{3}{5} h_0 \tau^{-1}$	& $\infty$ & $\infty$\\
\rule[-2ex]{0pt}{4ex}	&	$= \mathrm{par} \, \sigma_y^2(\tau) $	& $\approx 0.846 \,\mathrm{par} \, \sigma_y^2(\tau) $ & $=\mathrm{par} \, \sigma_y^2(\tau) $ & \\
\hline
\hline
\end{tabular}
\caption{Summary of the uncertainties $u$ associated to $\Pi$-, $\Lambda$-, and $\Omega$- weighted averages,  in comparison with the ADEV $(\sigma_y)$, modADEV ${(\rm mod} \, \sigma_y)$ and parADEV $({\rm par} \, \sigma_y)$ as a function of the averaging time $\tau$, and for different noise types $S_y(f)$, under the assumption $f_h \gg (2\pi\tau)^{-1}$. $\gamma \approx 0.577$ is the Euler-Mascheroni constant. PM stands for phase noise, FM for frequency noise.}
\label{tab:noise}
\end{table*}
To validate that the rapid fall-off of the modADEV and parADEV for white phase noise can be exploited by using $\Lambda$-averaging in conjunction with the associated modADEV /parADEV as an advantageous measure of uncertainty, the integrals given above were calculated. For completeness, we also provide the already known results for the ADEV. 

In the case of the ADEV, a high frequency cut-off $f_h \gg (2\pi \tau)^{-1}$ has to be introduced to ensure convergence of the integrals~\cite{all81}. In practice, high frequency noise is always cut off by the limited bandwidth of the components used or by deliberate low pass filtering. Furthermore, the ratio between $u_\Pi$ and $\sigma_y$ is independent of $f_h$  such that it does not affect the conclusions here.

We have listed the results for white phase, flicker phase, white frequency, flicker frequency, and random walk frequency noise in table~\ref{tab:noise}. Under the assumption $f_h \gg (2\pi\tau)^{-1}$, we have neglected a small sinusoidal contribution depending on $f_h$ in the corresponding entries. 
It can be seen that for the first three noise types the power law of the respective Allan variance type is identical to that of the corresponding uncertainty $u$. We draw two main conclusions: 
\begin{enumerate}
\item In the presence of dominating white or flicker phase noise one can indeed profit from the rejection of high frequency phase noise for the determination of the average $\overline{y}$ using $\Lambda$- or $\Omega$-weighting . 
\item The modADEV / parADEV can be employed to determine the uncertainty, provided that $\Lambda$- / $\Omega$-averaging and the appropriate correction factor listed in table~\ref{tab:noise} are applied. This is not conceptually different to using the ADEV to determine the uncertainty, but leads to tighter values of $u$ if white phase noise dominates.
\end{enumerate}
In practice, the question arises how to estimate the according Allan deviation for the determination of the uncertainty $u(\overline{y})$ of the average $\overline{y}$ taken over the entire available measurement length $T$. In the following we assume that the values of the Allan deviations at sampling times $\tau=T$ and below are known, e.g.\ from other measurements, as it was also presupposed in~\cite{lee10}.
Since the $\Lambda$-weighting function is nonzero over twice the averaging interval, the average value $\overline{y}$ and its uncertainty $u_\Lambda$ can be determined over an averaging interval of maximum length $T/2$. Instead, for the other weighting functions, which are restricted to exactly the averaging interval, the uncertainty can be determined for averaging time $T$.
\begin{table}
\centering
\begin{tabular}{l|c|c|c}
\hline
\hline
\rule[-1ex]{0pt}{4ex} 	&  White & Flicker & White \\ 
\rule[-2ex]{0pt}{4ex} Ratio	&  PM & PM & FM \\ 
\hline
\rule[-1ex]{0pt}{5ex}	$\frac{u_\Lambda^2(T/2)}{u_\Pi^2(T)}$	& $\frac{4}{f_h T}$ & $\frac{16\ln(2)}{3\ln(2\pi f_h T) -\ln(2) +3 \gamma}<1$ &	$\frac{4}{3}$\\
\rule[-1ex]{0pt}{5ex} $\frac{u_\Omega^2(T)}{u_\Pi^2(T)}$ &	$\frac{3}{f_h T}$ & $\frac{9}{3\ln(2\pi f_h T) -\ln(2) +3 \gamma}<1$ &	$\frac{6}{5}$ \\
\rule[-3ex]{0pt}{5ex} $\frac{u_\Omega^2(T)}{u_\Lambda^2(T/2)}$ & $\frac{3}{4}$ & $\frac{9}{16\ln(2)}\approx 0.812$ &	$\frac{9}{10}$\\
\hline
\hline
\end{tabular}
\caption{Ratios of the squared uncertainties obtained for $\Lambda$-, $\Pi$- and $\Omega$-averaging using the respective available averaging times that can be obtained from a given dataset of length $T$, under the assumption that $f_h \gg (2\pi\tau)^{-1}$.}
\label{tab:u_ratio}
\end{table}

The ratios $u_\Lambda^2(T/2)/u_\Pi^2(T)$, $u_\Omega^2(T)/u_\Pi^2(T)$ and $u_\Omega^2(T)/u_\Lambda^2(T/2)$ for the uncertainties determined by the described procedure are listed for typical noise types in table~\ref{tab:u_ratio}. Although for white and Flicker phase noise, some of the ratios depend on the $\Pi$-averaging cutoff-frequency $f_h$, $T$ is under practical conditions large enough to satisfy the conditions that $\Lambda$- and $\Omega$-averaging are advantageous over $\Pi$-averaging in these cases. However, this is not true if white frequency noise is dominating, where the application of $\Lambda$- or $\Omega$-type averaging is actually detrimental (ratio~$>1$), even though the direct comparison between ADEV, modADEV and parADEV in table~\ref{tab:noise} might not suggest that. This shows that it is important for the determination of the uncertainty to be aware of the difference between the classical variance and the two-sample variance such as the ADEV, modADEV or parADEV.

Comparing $\Omega$- with $\Lambda$-averaging reveals that $\Omega$- is in principle advantageous over $\Lambda$-averaging for all three noise types. However, in practice there are not yet many $\Omega$ frequency counters. For that reason, $\Omega$-averaging at present does not yet play an important role, although this may change in future with increasing digital signal processing capabilities. For the choice of the averaging method in post-processing-based average determination, $\Omega$-averaging is definitely a considerable option.
%
%
%
%
\section{Optimum averaging methods}
\label{sec:optimum}
The methodology described so far can be readily transferred to any other averaging method, e.g.\ describing counters with non-zero dead times~\cite{daw07,lee10} by adaptation of the weighting function. We now derive the optimum weighting functions for a given noise type by first recalling the general requirements and constraints imposed on the weighting function. (1) It leads to a minimum uncertainty, (2) must be confined to a finite averaging interval $\tau$ in the time domain, and (3) must be normalized to 1 for integration over this finite length interval. To minimize the uncertainty, i.e.\ the classical variance for a given power spectral density $S_y(f)$, we have to minimize the integral in equation~\ref{eq:u_FT}. For white phase noise $S_y(f)=h_2 f^2$, we find 
\begin{eqnarray}
u^2&=&h_2 \int_0^{\infty} f^2 \left|W(f)\right|^2 \, \rmd f \nonumber \\
&=&\frac{h_2}{4\pi^2}\int_0^{\infty}  \left|2\pi \rmi f W(f)\right|^2 \, \rmd f \nonumber\\
&=&\frac{h_2}{2\pi^2}\int_{-\tau/2}^{\tau/2} \left|\frac{\rmd w(t)}{\rmd t} \right|^2 \,\rmd t \nonumber\\
&=&\frac{h_2}{2\pi^2}\int_{-\tau/2}^{\tau/2} L(t,w,\dot{w})\,\rmd t,
\label{eq:L_int}
\end{eqnarray}
which has a functional form that can be minimized using the Euler-Lagrange formalism. We use the constraint 
\begin{equation}
\int_{-\tau/2}^{\tau/2} w(t) \,\rmd t = 1
\label{eq:w_constr}
\end{equation}
as Lagrange multiplier to define the function 
\begin{eqnarray}
L_\lambda &=& L(t,w,\dot{w})-\lambda \left (w - \frac{1}{\tau}\right) \nonumber\\
&=&\left|\dot{w}\right|^2-\lambda \left (w - \frac{1}{\tau}\right),
\label{eq:L}
\end{eqnarray}
and write down the corresponding Euler-Lagrange equation
\begin{equation}
\frac{\partial L_\lambda}{\partial w} -\frac{\rmd}{\rmd t} \frac{\partial L_\lambda}{\partial \dot{w}} = 0 \quad \Rightarrow \quad -\lambda-2\ddot{w} = 0.
\label{eq:EulerLagrange}
\end{equation}
Integration of equation~\ref{eq:EulerLagrange} using the boundary conditions $w(t\pm \tau/2)=0$ yields the $\Omega$-weighting function given by equation~\ref{eq:w_reg}. Hence, $\Omega$-averaging yields the minimum uncertainty when white phase noise dominates. Independently of and simultaneously to this work, it was realized by Vernotte et al. that the parADEV is the optimum choice to determine the frequency instability in presence of white phase noise~\cite{ver15a}.

For white frequency noise instead, we can write equation~\ref{eq:L_int} with $L=|w|^2$, and accordingly, equation~\ref{eq:EulerLagrange} reads
\begin{equation}
2 w-\lambda = 0 \quad \Rightarrow \quad w(t)=1/\tau \quad \mathrm{for}\quad 0<t\leq \tau.
\label{eq:EulerLagrange_WFM}
\end{equation}
Hence, the optimum weighting function in case of white frequency noise is the $\Pi$-weighting function.

%
%
%
%
%
%
\section{Correlated frequency data} \label{sec:cor}
It is instructive to investigate the subject from a slightly different point of view, that reflects a usual experimental situation: A counter with given averaging time $\tau$ and chosen weighting function $w_{\Pi / \Lambda}$ measures $N$ consecutive frequency averages $\overline{y}_w(l\tau)$, that shall in turn be averaged in a post-processing step to obtain the average $\overline{y}$ determined over the overall measurement time. The situation here is explicitly that the individual averaging windows are contiguously connected in the case of $\Pi$-averaging or even overlap with each other for $\Lambda$-averaging. This has direct consequences for the correlations between the individual averages $\overline{y}_w(l\tau)$. 

The uncertainties $u(\overline{y}_w)$ associated to the individual averages are known from the above analysis or via a direct estimate of the classical variance (equation~\ref{eq:classical}), and they can be assumed to be all equal in case of ergodic noise. The usual approach for the estimation of the combined uncertainty $u(\overline{y})$ of the average $\overline{y}$ of the dataset consisting of $N$ independent measurements is to multiply $u(\overline{y}_w)$ by $1/\sqrt{N}$.

However, in the general case correlations between the individual averages $\overline{y}_w(l\tau)$ have to be investigated and taken care of~\cite{wit09,lee10}. We will also see that for some situations, e.g.\ in which frequency data obtained with $\Lambda$-counters are subsequently $\Pi$-averaged, this approach leads to results more easily than an adaptation of $w(t)$ followed by the necessary integrations. The following method will also be used to optimize the averaging strategy for combined noise situations (section~\ref{sec:mixed}).

According to the ``propagation of uncertainties'' e.g.\ described in the GUM~\cite{gum08}, the combined uncertainty $u(f)$ of a function $f(x_l)$ is found by linearization of $f$ around the expectation values $\mu_l$ and depends on the uncertainties of the $N$ input quantities $x_l$ as 
\begin{eqnarray}
u^2(f) &=& \sum_{l,m=1}^N { \frac{\partial f}{\partial x_l} \frac{\partial f}{\partial x_m} u(x_l, x_m)} \nonumber \\
&=& \sum_{l=1}^{N}{ \left( \frac{\partial f}{\partial x_l}  \right)^2 u^2(x_l)}  \nonumber \\
 &&+2\sum_{l=1}^{N-1}\sum_{m=l+1}^N{\frac{\partial f}{\partial x_l} \frac{\partial f}{\partial x_m} u(x_l, x_m)},
\label{eq:u_f}
\end{eqnarray}
with the covariance between $x_l$ and $x_m$
\begin{equation}
u(x_l, x_m) = \left \langle (x_l - \mu_l)(x_m - \mu_m) \right \rangle.
\label{eq:covar}
\end{equation}

For the special case of $f = \overline{y} = \frac{1}{N} \sum_{l=1}^N \overline{y}_w(l\tau)$ being the $\Pi$-average of the $x_l=\overline{y}_w(l\tau)$ having identical uncertainties $u(\overline{y}_w)$ and a common expectation value $\mu$, we find~\cite{wit09,lee10}
\begin{eqnarray}
u^2(\overline{y}) &=& \frac{1}{N} u^2(\overline{y}_w) + \frac{2}{N^2} \sum^{N-1}_{l=1}{(N-l)C_{\overline{y}_w}(l \tau)} \nonumber \\
&=& \frac{u^2(\overline{y}_w)}{N} \left[ 1 + \frac{2}{N} \sum^{N-1}_{l=1}{(N-l)\frac{C_{\overline{y}_w}(l \tau)}{C_{\overline{y}_w}(0)}} \right],
\label{eq:u_f_5}
\end{eqnarray}
using the autocovariance function $C_{\overline{y}_w}(\tau)$ of $\overline{y}_w(t)-\mu$, i.e.\ of the fluctuations of the fractional frequency $y(t)$ convoluted with the weighting function $w(t)$. It is defined as
\begin{equation}
C_{\overline{y}_w}(\tau) = \left \langle \left[\overline{y}_w(t+\tau) - \mu\right]\left[\overline{y}_w(t) - \mu\right] \right \rangle.
\label{eq:rho}
\end{equation}
The Wiener-Khinchin theorem can be used to express the autocovariance function $C_{\overline{y}_w}(l \tau)$ as the Fourier transform of the relevant double-sided spectral density $s_{\overline{y}_w}(f)$ of $\overline{y}_w(t)-\mu$
\begin{eqnarray}
C_{\overline{y}_w}(l \tau) &=& \int^\infty_{-\infty} e^{2 \pi \rmi f l \tau} s_{\overline{y}_w}(f)\, \rmd f \nonumber \\
&=&\int^\infty_{-\infty} \cos(2 \pi f l \tau) s_{\overline{y}_w}(f)\, \rmd f \nonumber \\
&=&\int^\infty_0 \cos(2 \pi f l \tau) \left| W_{\Pi / \Lambda}(f) \right|^2 S_y(f)\, \rmd f,
\label{eq:R}
\end{eqnarray}
where in the second line we have used that the double-sided spectral density is even, because $\overline{y}_w(t)$ is real. In the third line, the double-sided spectral density $s_{\overline{y}_w}(f)$ has been substituted by the single-sided spectral density given by $S_y(f)$ weighted by the spectral weighting function $W_{\Pi / \Lambda}(f)$ and the integration interval has been changed accordingly. 

Table~\ref{tab:autocor} lists the values of the autocorrelation function defined by $C_{\overline{y}_w}(l\tau)/C_{\overline{y}_w}(0)$ and computed from equation~\ref{eq:R} for white frequency and white phase noise, $\Pi$- and $\Lambda$-averaging, and lag $l \tau$. We find that the autocorrelation  vanishes for $l \geq 2$ for the listed noise types (which is not the case for flicker phase noise) and that a $\Lambda$-counter introduces correlations due to overlapping averaging functions of the individual consecutive averaging processes, as can be seen in figure~\ref{fig:sensfct}.
\begin{table}
\centering
\begin{tabular}{l|c|c|c|c|c}
\hline
\hline 
\rule[-2ex]{0pt}{5ex}	& 					& \multicolumn{2}{c}{$\Pi$-counter}	& \multicolumn{2}{|c}{$\Lambda$-counter} \\ 
\rule[-2ex]{0pt}{5ex}Noise Type	  & $S_y(f)$	& $l = 1$				& $l\geq 2$					&  $l = 1$			& 	$l\geq 2$	 \\ 
\hline
\rule[-1ex]{0pt}{5ex}White PM   	& $h_2 f^2$				& $-\frac{1}{2}$		& 0 & $-\frac{1}{2} $	& 0\\
\rule[-2ex]{0pt}{6ex}White FM			& $h_0$						& 0			& 0 		& $\frac{1}{4}$	& 0			\\
\hline
\hline
\end{tabular}
\caption{Calculation of the autocorrelation function $C_{\overline{y}_w}(l\tau)/C_{\overline{y}_w}(0)$, using equation~\ref{eq:R} for different noise types, weighting functions and lag multipliers $l$.}
\label{tab:autocor}
\end{table}
\begin{table}
\centering
\begin{tabular}{l|c|c}
\hline
\hline 
\rule[-2ex]{0pt}{5ex}Noise Type		& $\Pi$-counted $\overline{y}_w$	& $\Lambda$-counted $\overline{y}_w$ \\ 
\hline
\rule[-1ex]{0pt}{5ex}White PM   	& $\frac{u_{\Pi}^2}{N^2}$	& $\frac{u_{\Lambda}^2}{N^2}$\\
\rule[-2ex]{0pt}{6ex}White FM			& $\frac{u_{\Pi}^2}{N}$	& $\frac{3N-1}{2N^2} u_{\Lambda}^2 \approx \frac{3}{2} \frac{u_{\Lambda}^2}{N} $\\
\hline
\hline
\end{tabular}
\caption{Overall average $u^2(\overline{y})$ for a set of $N$ $\Pi$- or $\Lambda$-counted averages $\overline{y}_w$ under the assumption of white phase or white frequency noise.}
\label{tab:u_scale}
\end{table}

Inserting the results in table~\ref{tab:autocor} into equation~\ref{eq:u_f_5}, we can determine the dependence of $u^2(\overline{y})$ on $N$. The respective results are listed in table~\ref{tab:u_scale}. The dependence of $u^2(\overline{y})$ on $N$ expected for white phase and white frequency noise is retrieved for $\Pi$-counted frequency averages $\overline{y}_w$~\cite{lee10, wit09} and also found for the case of $\Lambda$-counters. We have thus built up a coherent theoretical framework that can be employed to find further optimized data analysis strategies in the presence of combined noise types. 
%
%
%
%
%
%
\section{Optimization for a transition between noise types}\label{sec:mixed}

\begin{figure*}\
\centerline{\includegraphics[width=2\columnwidth]{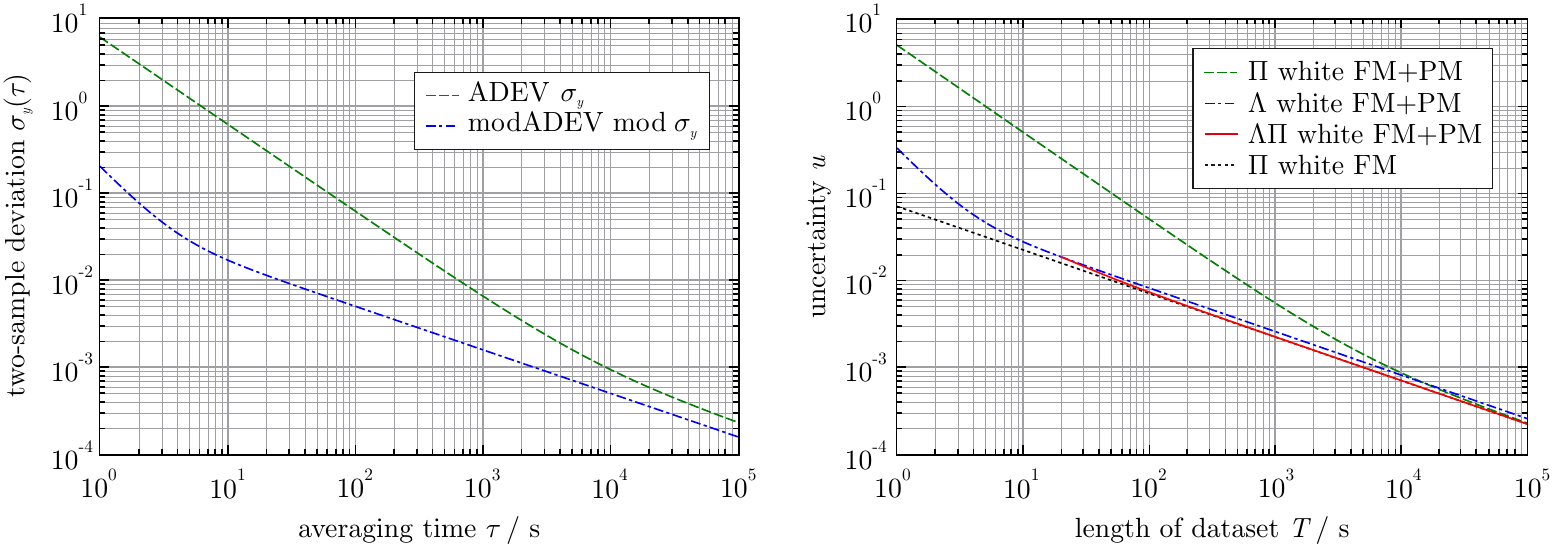}}
\caption{(left) ADEV and modADEV for data with a power spectral density $S_y(f)$ of combined white frequency and white phase noise. Parameters are chosen such that the phase noise dominates for a significant part of the relevant measurement interval ($h_0=0.01$~s, $h_2 = 10 \,\mathrm{s}^3$, $f_h = 50$~Hz, $\tau_0 = 1$~ms). For a discussion of the $\tau_0$ parameter see section~\ref{subsec:counters}.\\
(right) Calculation of $u$ as a function of available data for the same noise spectrum as left using pure $\Pi$-averaging (equation~\ref{eq:sens_av_pi}, dashed green line), pure $\Lambda$-averaging (equation~\ref{eq:sens_av_lam}, dashed-dotted blue line), and $\Lambda$-averaging up to a given $\tau^\prime=10$~s followed by $\Pi$-averaging of 10~s $\Lambda$-counted values (solid red line). Using the optimized averaging approach, the white frequency noise limit (dotted black line) is approached at much shorter averaging times than for $\Pi$-averaging.}
\label{fig:allans}
\end{figure*}

We have shown in section~\ref{sec:modADEV} that $\Lambda$-type averaging is advantageous over $\Pi$-averaging in the presence of phase noise to derive measurement results with small uncertainty $u$ and that the modADEV can be used to determine $u$. Often, more than one type of noise is relevant in measurements. The combination of white phase and white frequency noise is a common example. Depending on the averaging time, one or the other noise type will dominate. Thus, according to table~\ref{tab:noise}, using the advantageous uncertainty measure demands different average weighting functions in the $\tau$ intervals differing in the dominating noise types.

We will now develop an optimization strategy of mixed $\Lambda$- and $\Pi$-averaging to account for such a situation. For illustration it is helpful to consider a measurement situation that leads to an ADEV or modADEV as shown in figure~\ref{fig:allans} (left). A comparison of optical clocks via a long fibre frequency link~\cite{pre12} or via a satellite link can lead to similar plots.

Calculating and plotting  $u$ (figure~\ref{fig:allans}~(right)) according to section~\ref{sec:modADEV} for $\Pi$- and $\Lambda$-averaging illustrates the findings of tables~\ref{tab:noise} and \ref{tab:u_ratio}. Due to the linearity of equation~\ref{eq:u_FT} in $S_y(f)$, the uncertainty $u(\overline{y}_w)$ is the square root of the sum of squares of the uncertainties stemming from the individual noise contributions.

Switching from $\Lambda$- to $\Pi$-averaging at an averaging time $\tau^\prime$ allows to unite the advantageous behaviour for both averaging approaches for the respective noise type (figure~\ref{fig:allans} solid red line). Mathematically, this averaging approach can be described by an adapted weighting function $w_{\Lambda \Pi}$ or more easily by the approach outlined in section~\ref{sec:cor}. We use $\Lambda$-averaged $\overline{y}_w$ with $\tau = \tau^\prime = 10$~s and average the $N = T/\tau^\prime -1$ data points $\overline{y}_w$ that can be determined from an overall measurement length $T$. Explicit calculation of $u^2(\overline{y})$ in equation~\ref{eq:u_f_5} for combined noise using equations~\ref{eq:u_FT} and~\ref{eq:R} reveals that $u(\overline{y})$ again is the square root of the sum of squares of the uncertainties due to the individual noise contributions.
%
%
%
%
%
%
\section{Practical implications} \label{sec:practical}
The methodology described in the previous sections can be applied to frequency measurements by direct clock comparisons as well as to clock comparisons via some kind of link, e.g. a fibre or satellite link. The link-based comparisons are essentially also frequency measurements, but the drift and fluctuations of the link during a measurement enter the results as well.

In the following, we discuss some more practical aspects occurring with real-world equipment used for frequency measurements, which are modern high-resolution frequency counters as well as satellite timing and ranging equipment (SATRE) modems. These aspects are closely related to further post-processing analysis of raw data generated by this equipment.

\subsection{State-of-the-art frequency counters}
\label{subsec:counters}
In the past decades, frequency counter technology evolved~\cite{kal04,rub05, daw07} from conventional counting of integer periods during a certain gate time, via reciprocal and interpolating reciprocal counting, which adapts the gate time to an integer multiple of periods, to modern continuous phase recording based frequency counters, which are also referred to as continuous time-stamping frequency counters~\cite{kra01, kra04a, joh05}. 

Essentially, the latter counter type measures frequencies obtained by dead-time-free $\Pi$-weighted frequency averaging over progressive time intervals of length $\tau_0$ referenced to a signal with frequency $f_\mathrm{ref}$, i.e.\
\begin{equation}
\tau_0=\frac{k}{f_\mathrm{ref}},
\label{eq:tau0}
\end{equation}
where $k$ is a fixed factor. The $\Pi$-weighted frequency average $f_\Pi$ is achieved by a determination of the fractional number of cycles $C_n$ of the signal under test that fits into the $n$-th averaging time interval, divided by the interval length $\tau_0$. $C_n$ is the phase difference accumulated over the averaging time interval $\tau_0$ in units of $2\pi$.
\begin{equation}
f_\Pi = \frac{\phi(t_n+\tau)-\phi(t_n)}{2 \pi \tau_0} = \frac{C_n}{\tau_0} = \frac{C_n}{k}f_\mathrm{ref},
\label{eq:cycles}
\end{equation}
which shows that the phase recorder actually measures the frequency ratio between the signal under test and the reference signal. Furthermore, due to the absence of dead times between the averaging intervals, the phase values $\phi(t_n)=\sum_n C_n/2\pi$ are exactly given by the integration over the frequency values. The fractional number of cycles is determined by the combination of two techniques measuring the integer and the non-integer part separately. The integer part is determined with a continuously running counter, whose value is read out on-the-fly every $\tau_0$. This strobed on-the-fly readout leads to the absence of dead times between the consecutive averaging intervals. The non-integer part is measured with analogue interpolation electronics, e.g.\ based on charging a capacitor at a constant current~\cite{kra04a}.

More complex frequency averaging weighting functions for longer averaging times $\tau\ge\tau_0$ can be synthesized by proper addition and normalization of short $\Pi$-weighting functions at $w_\Pi(t+n\tau_0)$ of length $\tau_0$, e.g. in order to approximate a $\Lambda$ weighting function~\cite{rub05}. Alternatively, this can be done by according processing of the time series of the phase values as, e.g. by a linear regression for $\Omega$-averaging.

Due to the granularity on the $\tau_0$ grid of the synthesized weighting function, the two-sample deviation associated to the averaging method (e.g.\ the modADEV associated to synthesized $\Lambda$-weighted frequency averaging) coincides at $\tau_0$ with the Allan deviation $\sigma(\tau_0)$, i.e.\ with the two-sample variance associated to $\Pi$-weighted frequency averaging (see~\cite{rub05} for the case of a synthesized $\Lambda$ weighting function). This value $\sigma(\tau_0)$ strongly depends on the cutoff-frequency $f_h$ for the $\Pi$-averaging. Hence, in order to make Allan deviation plots generated from different experiments comparable, it is important to provide the value of $f_h$ if an Allan deviation is derived from experimental data measured with $\Pi$-counters, and to provide both the values $f_h$ and $\tau_0$ if another two-sample deviation (e.g.\ the modADEV) variant is determined using a ``synthesized'' weighting function for averaging. 

We consider as an example the K+K GmbH FXE frequency counter~\cite{kra04a}, which is a synchronously sampling multichannel progressive phase recording frequency counter nowadays used in many frequency metrology labs. It implements phase samples on a $\tau_0\approx 1$~ms grid. This value should not be confused with the resolution of the phase interpolation, which corresponds to the resolution at which $\tau_0$ is determined and has a value of 12.2~ps for the FXE. The FXE allows a large cutoff frequency $f_h$ of several tens of Megahertz. Thus, $f_h$ is usually determined by deliberate filtering of the signal before applying it to the input of the FXE. The FXE can save the raw phase values $\phi(t_n)$ time-stamped on the $\tau_0=1$~ms grid to a file on a computer. Alternatively, the internal microprocessor of the FXE can perform an on-the-fly $\Pi$- or $\Lambda$-averaging for averaging times choosable between 1~ms and 30~s and save them to a file on a computer for further post-processing. The synchronously sampled, in principle unlimited number of parallel counter channels of the FXE allows to measure differential phases or frequencies between the input signals. Fluctuations of the reference signal applied to the FXE are thus eliminated by a very high common mode rejection.
\subsection{SATRE modems}
As an alternative method to determine a coherent time series of phase values $\overline{\phi}_n$ on a $\tau_0$ grid, the sampled output of a phase comparator can be used. This is usually done in SATRE modems for two-way satellite time and frequency transfer (TWSTFT)~\cite{sch99c}. It is straightforward that the output of the phase comparator depends on its nonlinearity, which may especially be critical for an analogue phase comparator. Nevertheless, noise can be filtered out either due to limited bandwidth of the delay locked loop employed for code locking or with a narrow low-pass filter behind the phase comparator, and in order to obtain the phase samples, the temporal phase can be averaged according to 
\begin{equation}
	\overline{\phi}(t_n)=\frac{1}{\tau_0}\int_{t_n-\tau}^{t_n+\tau} \phi(t)\, \rmd t,
\end{equation}
which yields $\Lambda$-weighted fractional frequency averages given by
\begin{equation}
	\overline{y}_{\Lambda}(t_n)=\frac{\overline{\phi}(t_{n+1})-\overline{\phi}(t_n)}{2\pi\tau_0}.
\end{equation}
The values of the $\overline{y}(t_n)$ time series can be read out to a computer and written to a file as a time-stamped list on a $\tau_0=1$~s grid. Again, these data can be further processed in order to obtain frequency averages with synthesized weighting functions. For example to approximate $\Pi$-weighted frequency averages by short $\Lambda$ averages, we have to determine the difference between the last and first phase value of the $\phi(t_n)$ time series and divide by $\tau$. For a $\Lambda$-averaged frequency over the interval $\tau$, we have to subtract the equally weighted average over the $\phi(t_n)$ in the interval $0 \dots \tau$ and over those in the interval $-\tau \dots  0$ from each other and divide by $\tau$. Please note that for $\Lambda$-averaging, there is no granularity involved in this case, because the elementary averages at $\tau_0$ are of the same kind. For linear regression weighting, the average angular frequency is given by the slope of the linear regression line fitted to the time series $\phi(t_n)$ within the interval $\tau$. Regression weighting is particularly advantageous for clock comparisons via links with SATRE-modems, because it is robust against gaps in the phase data, which are always present using real satellite links~\cite{ses08}. 
%
%
%
%
%
\section{Conclusion} \label{sec:con}

This paper provides the necessary mathematical foundation for the determination of the statistical uncertainty of a frequency measurement using the modified Allan deviation and other variants of the Allan deviation. When white phase noise is the dominating noise type, the steeper power law dependence on averaging time of the modified Allan deviation and of the parabolic Allan deviation with respect to the Allan deviation can indeed be exploited to achieve a tighter uncertainty at identical averaging times. This has direct benefits like, e.g., shorter measurement times required for comparison of optical clock frequencies using satellite links. For optical clock comparisons via fibre links \cite{dro13, rau14, rau15}, we see that $\Lambda$-averaging will allow to reduce the link instability to below the state-of-the-art instability of optical lattice clocks of $2 \times 10^{-16}/\sqrt{\tau / {\rm s}}$ after a few 10~s. For pure $\Pi$-averaging however, several hours of averaging would be required. Keeping in mind that phase coherence has to be maintained over this interval, we find strongly relaxed requirements on the operational stability of the fibre links. 

Conversion factors between the modified Allan deviation and the  uncertainty of the relative frequency have been found for several typical noise types. The presented formalism is not limited to the discussed noise types and the variants of the Allan deviation, but rather applicable to any two-sample variance associated to an averaging weighting function. 

We derived that for predominant white phase noise, linear regression  weighting ($\Omega$ frequency averaging) yields a minimum uncertainty, as does $\Pi$ averaging in case of white frequency noise.
Furthermore, we have shown that employing the modified Allan deviation for the determination of the uncertainty can also have benefits even with white phase noise on a white frequency background. A formalism to use an advantageous uncertainty measure in case of combined noise contributions has been presented, which includes the practically important situation of white phase noise combined with white frequency noise. 

\ack
This work was performed within the framework of the Centre of Quantum Engineering and Space-Time Research (QUEST). Partial funding from the German Research Foundation DFG within the geo-Q collaborative research center CRC~1128 and reasearch training group RTG~1729 is acknowledged, as well as funding within the ITOC and QESOCAS projects in the European Metrology Research Programme EMRP. The EMRP is jointly funded by the EMRP participating countries within EURAMET and the European Union. 

The authors are grateful for useful discussions with Burghard Lipphardt and Sebastian Raupach.

\section*{References}
\bibliographystyle{prsty}
\bibliography{texbi431}

\end{document}